\documentclass[aps,prd,a4paper,twocolumn,showpacs,showkeys,preprintnumbers,amsmath,amssymb,nofootinbib,usenames,dvipsnames]{revtex4-2}
\usepackage{graphicx}
\usepackage{bm}
\usepackage{dcolumn}
\usepackage{color}
\usepackage{multirow}

\usepackage[titletoc]{appendix}

\usepackage{mathrsfs} 
\usepackage{txfonts} 

\newcommand{\bea}{\begin{eqnarray}}
\newcommand{\eea}{\end{eqnarray}}

\def\beq{\begin{equation}}
\def\eeq{\end{equation}}

\newcommand{\rd}{\mathrm{d}}  

\def\al{\alpha}
\def\be{\beta}
\def\ga{\gamma}

\def\de{\delta}
\def\De{\Delta}

\def\th{\theta}

\def\rh{\rho}
\def\si{\sigma}

\def\ta{\tau}
\def\up{\upsilon}

\begin{document}

\title{Toward regular black holes in sixth-derivative gravity
\vspace{0.2cm}}

\author{Breno L. Giacchini}
\email{breno.giacchini@matfyz.cuni.cz}
\affiliation{
{\small Institute of Theoretical Physics, Faculty of Mathematics and Physics, Charles University, V Hole{\v s}ovi{\v c}k{\'a}ch 2, 180 00 Prague 8, Czech Republic}
}

\author{Ivan Kol\'a\v{r}}
\email{ivan.kolar@matfyz.cuni.cz}
\affiliation{
{\small Institute of Theoretical Physics, Faculty of Mathematics and Physics, Charles University, V Hole{\v s}ovi{\v c}k{\'a}ch 2, 180 00 Prague 8, Czech Republic}
}


\begin{abstract} 
We study spherically symmetric static solutions of the most general sixth-derivative gravity using series expansions. Specifically, we prove that the only solutions of the complete theory (i.e., with generic coupling constants) that possess a Frobenius expansion around the origin, $r=0$, are necessarily regular. When restricted to specific branches of theories (i.e., imposing particular constraints on the coupling constants), families of potentially singular solutions emerge. By expanding around $r=r_0 \neq 0$, we identify solutions with black hole horizons. Finally, we argue that, unlike in fourth-derivative gravity, the conditions $R=0$ and $g_{tt}g_{rr}=-1$ are too restrictive for sixth-derivative gravity solutions.
\end{abstract}

\maketitle
\noindent


\section{Introduction}

In the last decade there has been a renewed interest in the important yet ambiguous role of higher derivatives in gravity. The perturbative quantization of Einstein gravity and even the quantization of matter fields on a curved background require the introduction of higher-derivative terms in the gravitational sector to renormalize loop divergences~\cite{UtDW,hove,GorSag12}. Such terms also appear in the low-energy regime of string theory~\cite{Candelas:1985en,Accioly:2016qeb}. 
Alternatively, by considering higher-derivative terms at the fundamental level, one can formulate renormalizable and superrenormalizable models of quantum gravity~\cite{Stelle77,AsoreyLopezShapiro}. 
This comes with the drawback of having ghostlike particles in the spectrum, traditionally associated with violation of unitarity and other instabilities. Recently, however, considerable effort has been made to analyze scenarios in which the effect of ghosts can be controlled (at classical and quantum levels) and unitarity can be recovered~\cite{Bender:2007wu,Salvio:2015gsi,Deffayet:2021nnt,Anselmi:2017ygm,Donoghue:2019fcb,ModestoShapiro16}.
Some of these constructions depend on more complicated actions, with metric derivatives higher than fourth~\cite{AsoreyLopezShapiro,ModestoShapiro16} or in the form of nonlocal operators~\cite{Krasnikov,Kuzmin,Modesto12,SiegelEtAl3}.

Among the main issues that a quantum theory of gravity is expected to address is the resolution of the spacetime singularities that occur in general relativity. In the absence of a satisfactory theory and an underlying regularization mechanism, phenomenological models of quantum-corrected regular black holes abound in the literature. 
The question of whether higher derivatives could 
solve the problem 
has been considered in different contexts~\cite{Modesto12,SiegelEtAl3,Tseytlin:1995uq,Frolov:Poly,SiegelEtAl3,Buoninfante:2018xiw,Frolov:2023jvi,Kolar:2023gqi,Kolar:2021uiu,dePaulaNetto:2023vtg,BreTib1,Nos6der}, but owing to the complicated equations of motion and the humongous amount of possible higher-derivative correction terms,
such studies most often involve a linearization of field equations.
In this simplified setup it was shown that there is a significant difference between 
theories defined by actions with four and with \textit{more} than four metric derivatives. For instance, the latter models have a regular nonrelativistic limit when coupled to normal matter, while the former still displays curvature singularities~\cite{BreTib1,Nos6der}.

Nevertheless, results concerning exact solutions of higher-derivative gravities are nearly absent in the literature. An important exception is the case of fourth-derivative gravity, for which several families of exact solutions are known 
(see, e.g.,~\cite{Stelle78,Holdom:2002xy,Lu:2012xu,Lu:2015cqa,Stelle15PRD,Cai:2015fia,Feng:2015sbw,Holdom:2016nek,Holdom:2022zzo,Kokkotas:2017zwt,Goldstein:2017rxn,Podolsky:2019gro,Svarc:2018coe,Salvio:2019llz,Bonanno:2019rsq,Daas:2022iid,Silveravalle:2022wij} and references therein).
It turns out that this model might not offer a resolution of the singularity problem at the classical level. Indeed, it admits singular static spherically symmetric solutions --- and the asymptotically flat solutions that couple to normal ghost-free matter appear to contain a naked singularity~\cite{Stelle15PRD}.

A suggestion that the situation could be different for gravity models with more derivatives was provided in the work~\cite{Holdom:2002xy}, 
which reported finding regular solutions, 
but a systematic study of the models and their solutions remained open. Recently, exact solutions for the Einstein gravity augmented by the six-derivative term ${C^{\mu\nu}}_{\al\be} {C^{\al\be}}_{\rh\si} {C^{\rh\si}}_{\mu\nu}$ have been obtained~\cite{Daas:2023axu}. While this term can be regarded as a two-loop correction based on the quantization of general relativity~\cite{GorSag12}, it is not the only one at this perturbative order. Moreover, taken alone it does not shed light on the features of the solutions of superrenormalizable gravity models, nor does it offer an immediate insight into the singularity problem, as divergent solutions also seem to exist~\cite{Praha2}. 

Exact solutions are also known for the Einsteinian cubic gravity and higher-dimensional quasitopological theories of gravity~\cite{Hennigar:2016gkm,Bueno:2016lrh,Hennigar:2017ego,Hennigar:2017ego,Bueno:2017sui}. Both classes of models are based on actions with very specific combinations of higher-derivative terms in order to yield second-order field equations for static spherically symmetric metrics. However, owing to the uniqueness of such constructions, their solutions ought to be the exception rather than the rule for higher-derivative gravities. For instance, regular black holes have been recently obtained in this framework but only in higher-dimensional theories with an infinite tower of higher-derivative terms~\cite{Bueno:2024dgm,DiFilippo:2024mwm}.

Our goal in this short paper is to present  results regarding the static spherically symmetric vacuum solutions in gravity models defined by actions with up to six derivatives of the metric, especially concerning the occurrence of regular spacetime configurations. The basic assumption is that such higher-derivative action is the relevant one at some energy scale, regardless of whether it is a fundamental theory or an emergent one. We shall not attempt to discuss the complicated problem of ghosts and stability of the solutions, as this lies beyond the scope of this work.

The gravity model we study is the most general extension of the Einstein--Hilbert action that includes terms with four and six derivatives of the metric, namely,
\bea
\label{mostgeneralaction}
S & = & \int \rd^4 x \sqrt{-g} \Big[  \al  R  
+ \be_1 R^2 +  \be_2 R_{\mu\nu}^2 
+ \ga_1 R \Box R 
+ \ga_2 R_{\mu\nu} \Box R^{\mu\nu} 
\nonumber
\\
&& + \, \ga_3 R^3 
+ \ga_4 R R_{\mu\nu} R^{\mu\nu} 
+ \ga_5 R_{\mu\nu} R^\mu{}_\rh R^{\rh \nu}
+ \ga_6 R_{\mu\nu} R_{\rh\si} R^{\mu\rh\nu\si} 
\nonumber
\\
&&
+ \, \ga_7 R R_{\mu\nu\rh\si} R^{\mu\nu\rh\si}
+ \, \ga_8 R_{\mu\nu\rh\si} R^{\mu\nu\ta\up} R^{\rh\si}{}_{\ta\up} 
\Big] 
,
\eea 
where the constants $\al$, $\be_{1,2}$, and $\ga_{1,\ldots,8}$ are, respectively, the coefficients of the terms with a total number of 2, 4, and 6 derivatives. 
Any other four- or six-derivative term can be cast as a combination of the terms in~\eqref{mostgeneralaction} and boundary or topological terms (that do not contribute to the equations of motion)~\cite{Decanini:2007gj,Harvey:1995}. 

The variation of the above action with respect to the metric yields the field equations
\beq
\label{EoM}
H_{\mu\nu} \equiv \frac{1}{\sqrt{-g}}  \frac{\de S}{\de g^{\mu\nu}} = 0,
\eeq
which we calculated using the package \textsc{xAct}~\cite{xAct,Brizuela:2008ra,xCoba} for \textit{Mathematica}~\cite{Mathematica}.
For a generic metric in the standard spherically symmetric coordinates,
\beq
\label{metric}
\rd s^2 = - B(r) \rd t^2 + A(r) \rd r^2 + r^2 \left( \rd \th^2 + \sin^2 \th \rd \phi^2 \right) 
,
\eeq
the field equations assume a diagonal form. Moreover, the generalized Bianchi identity $\nabla^\nu H_{\mu\nu} = 0$ acts as a constraint, and we end up with only two independent equations that can be taken to be $H_{tt}=0$ and $H_{rr}=0$.
Together, they form a system of coupled differential equations that is of sixth order for $B(r)$ and fifth order for $A(r)$. The terms of highest differential order are originated from the structures proportional to $\ga_1$ and $\ga_2$ in~\eqref{mostgeneralaction}, as these are the ones that contain the largest number of derivatives acting on a single metric component.

Solutions of the field equations around a certain point $r = r_0$ can be obtained by assuming that the functions $A(r)$ and $B(r)$ are represented by Frobenius series. For expansions around $r=0$, we use the ansatz
\beq
\label{FrobeniusAB}
A(r) = r^s \sum_{n=0}^\infty a_{n} r^n , \qquad  \frac{B(r)}{b_0} = r^t \left( 1 + \sum_{n=1}^\infty b_{n} r^n \right)  ,
\eeq
with $s,t \in \mathbb{R}$ yet to be determined, while around a generic point $r_0 \neq 0$ we use the more convenient representation in terms of $F(r) \equiv 1/A(r)$,
\beq
\label{FrobeniusFB}
F(r) = \De^w \sum_{n=0}^\infty f_{n} \De^n , \quad  \frac{B(r)}{b_0} =  \De^t \left( 1 + \sum_{n=1}^\infty b_{n} \De^n \right)  ,
\quad \De \equiv r - r_0
\eeq
with $w,t \in \mathbb{R}$ to be determined. 
We assume that $a_0,b_0,f_0 \neq 0$, so that the powers $s$, $t$, and $w$ define the leading terms of the series.
The corresponding expansion of the field equations at the lowest order comprises the system of indicial equations, for they determine the admissible values of $s$ (or $w$) and $t$. This is a first constraint on the space of possible solutions. 
After fixing this pair of parameters, a solution might be obtained by solving the field equations order by order.
We refer to the families of solutions by the pair of indexes $(s,t)_0$ and $(w,t)_{r_0}$, with the subscript label indicating whether this is a solution around $r=0$ or $r=r_0\neq 0$.

From the differential order of the field equations, one might expect that solutions could have up to 11 free parameters [this counting includes the parameter $b_0$, which is not physical for it corresponds to the time rescaling freedom of the metric~\eqref{metric}]. Although the field equations form a nonlinear system, the reasoning based on the differential order seems to work in quadratic gravity~\cite{Stelle15PRD}. As we show below, all the solutions we found satisfy this upper bound.

\section{Solutions around $r=0$}

The expanded field equations at the lowest order using the ansatz~\eqref{FrobeniusAB} have the general structure 
\beq
\begin{split}
\frac{H_{tt}(r)}{B(r)} &= r^{p(s)} \sum_{i=1}^8 \ga_i \, g_i(s,t,a_0)   + \ldots  = 0,
\\
H_{rr}(r) &= r^{q(s)} \sum_{i=1}^8 \ga_i \, h_i(s,t,a_0) + \ldots  = 0,
\end{split}
\label{EoMLow}
\eeq
where the powers $p$ and $q$ depend on $s$, the coefficients $g_i$ and $h_i$ only depend on $s$, $t$, and $a_0$, and the ellipsis denote terms of higher order in $r$. The explicit expressions for all these quantities are presented in the Appendix~\ref{App}. Notice that, to the lowest order,~\eqref{EoMLow} only receives contributions from the six-derivative terms in the action~\eqref{mostgeneralaction}.

The requirement that the field equations are solved for arbitrary values of the parameters $\ga_{1,\ldots,8}$, results in the system of indicial equations
\beq
\label{indicial}
g_i(s,t,a_0) = h_i(s,t,a_0) =  0, \quad  \forall \, i=1,\ldots,8 .
\eeq
A detailed consideration of this system is carried out in the Appendix~\ref{App}, where we also prove that its only solution is
\beq
s=t=0, \qquad a_0 = 1.
\eeq 
In other words, there is only one family of solutions of the type~\eqref{FrobeniusAB} around $r=0$, with indicial structure $(0,0)_0$.

After solving the equations $H_{tt}=0$ and $H_{rr}=0$ up to nine orders in $r$, we are convinced that the solutions in this family are characterized by six parameters, which can be taken to be $a_2$, $a_3$ (or $b_3$), $a_4$, $b_0$, $b_2$, and $b_4$. Among these, only five parameters are physical, for $b_0$ corresponds to the time rescaling freedom of the metric.
The free parameters appeared within the first five orders of the expansion; beyond this, at each new order we found two equations for two new parameters. The general structure of the solution is
\begin{eqnarray}
A(r) & = &  1 
+ a_2 r^2 
+ a_3  r^3
+ a_4 r^4 
+ \frac{(a_3+b_3)\bar{a}_5}{\ga_2 (3 \ga_1 + \ga_2) } r^5
+  O(r^6),
\nonumber
\\ \label{Sol00Geral}
\\
\frac{B(r)}{b_0} & =& 1 
+ b_2 r^2
+ b_3 r^3
+ b_4 r^4
+ \frac{(a_3+b_3)\bar{b}_5}{\ga_2 (3 \ga_1 + \ga_2)} r^5 + O(r^6),
\nonumber
\end{eqnarray}
where the parameters $a_3$ and $b_3$ are related through
\beq
(8  \ga_1 + 3  \ga_2 ) a_3 = 3  (4 \ga_1 + \ga_2 ) b_3,
\eeq
and the quantities $\bar{a}_5$ and $\bar{b}_5$ depend polynomially on $a_2$ and $b_2$ and on the parameters of the model. The coefficients of the terms $O(r^6)$ are determined from the lower-order ones.

A remarkable feature of these solutions is that the geometry is regular at $r=0$ in the sense that all the curvature invariants constructed by contracting an arbitrary number of Riemann and metric tensors are bounded. This happens because the solution in Eq.~\eqref{Sol00Geral} satisfies $a_0 = 1$ and $a_1 = b_1 = 0$. 
In particular, for the Kretschmann and Ricci scalars we have
\begin{align}
R_{\mu\nu\al\be} R^{\mu\nu\al\be} & \underset{r \to 0}{=} 12 \left( a_2^2 + b_2^2 \right) + O(r) ,
\\
R & \underset{r \to 0}{=} 6 \left(  a_2 - b_2 \right)  + O(r) .
\end{align}

It is also interesting to notice that the solution~\eqref{Sol00Geral} explicitly requires $\ga_2 (3 \ga_1 + \ga_2) \neq 0$. This is precisely the condition for the model~\eqref{mostgeneralaction} to have sixth-order derivatives in both the spin-2 and spin-0 sectors. If this condition is not satisfied, the structure of the solution family $(0,0)_0$ can be different from the one of the most general model. For instance, if $\ga_1 = \ga_2 = 0$ the field equations contain at most four derivatives acting on a single metric function, and one might expect a reduction of the number of free parameters in a solution. In Sec.~\ref{Sec.Ren} we shall discuss some important cases of incomplete sixth-derivative models, i.e., in which some coefficients $\ga_i$ are null or assume other particular values.

\section{Solutions around $r=r_0\neq 0$}

In contrast to the solutions around $r=0$, we have identified several families of solutions around a finite point $r=r_0\neq 0$. This diversity is indicative of the various possibilities for the point $r_0$, which can be, e.g., a horizon or a generic point, depending on the indicial structure of the solution~\eqref{FrobeniusFB}.

The solutions around a generic point are in the class $(0,0)_{r_0}$. After solving the field equations up to three orders in $r-r_0$, we are convinced that such solutions are characterized by 11 free parameters ($f_0,\ldots, f_4$, $b_0,\ldots, b_4$, and $r_0$), which appear already at first order. This is precisely the maximal number of free parameters (including the nonphysical parameter $b_0$) that we expected for a solution of the type~\eqref{FrobeniusFB}, taking into account the differential order of the field equations. The situation here is analogous to the fourth-derivative gravity, where the family of solutions $(0,0)_{r_0}$ around a generic point also has the maximal number of free parameters~\cite{Stelle15PRD}. This counting of free parameters is confirmed by the analysis using the metric written in conformal-to-Kundt form~\cite{Praha2}.

Expansions around a horizon $r=r_0$ correspond to the family $(1,1)_{r_0}$. Similar analysis reveals that these solutions are characterized by 5 physical parameters (which can be taken to be $f_0$, $f_1$, $b_1$, $b_2$, and $r_0$) and their existence allows us to conclude that there are black holes in sixth-derivative gravity. The general structure of the solution is
\begin{eqnarray}
F(r) & = &  
  f_0 \De
+ f_1 \De^2 
+ f_2  \De^3
+ f_3  \De^4
+  O(\De^5),
\nonumber
\\
B(r) & =&  b_0 \left( 
 \De
+ b_1 \De^2 
+ b_2  \De^3
+ b_3  \De^4
\right)  
+  O(\De^5),
\label{Sol11r0}
\end{eqnarray}
where 
four parameters among $f_0$, $f_1$, $f_2$, $b_1$, $b_2$ are independent, $b_0$ represents the residual gauge freedom,  and the subsequent parameters $f_n$ and $b_n$ for $n\in\lbrace 3,4,\ldots\rbrace$ can be expressed as functions of those and of $r_0$. Different from the solution $(0,0)_0$, in this case we refrain from displaying the structure of the first terms of the solution because the expressions are considerably longer. Nevertheless, Eq.~\eqref{Sol11r0} is enough to verify that the solutions of type $(1,1)_{r_0}$ have regular curvature scalars at the horizon $r=r_0$. In fact, the Kretschmann and Ricci scalars are bounded,
\begin{align}
R_{\mu\nu\al\be} R^{\mu\nu\al\be}  & \underset{r \to r_0}{=}  \frac{ 16 + 16 f_0^2 r_0^2 +  (3 b_1 f_0+f_1)^2  r_0^4 }{4 r_0^4} + O(\De) ,
\\
R  & \underset{r \to r_0}{=}  \frac{4- [f_0 ( 8 + 3 b_1 r_0 ) + f_1 r_0 ]r_0}{2 r_0^2} + O(\De) .
\end{align}

Since asymptotically flat regular black holes must have an even number of horizons, one expects also to encounter families of solutions expanded around a double (extreme) horizon. The search for such solutions is more complicated precisely because of their extremal nature, which makes them completely determined by the parameters of the model (with the only free parameters being the gauge one). More details about these and other solutions around $r=r_0\neq 0$ will be provided in a separate work~\cite{Praha2}.

\section{Solutions with $R=0$ and with $g_{tt} g_{rr} = -1$}

There are two classes of solutions that, although prominent in general relativity and quadratic gravity, do not seem to be very relevant for a generic sixth-derivative gravity model.

In fourth-derivative gravity, metrics with vanishing Ricci scalar ($R=0$) played an important role in the identification of certain classes of asymptotically flat solutions~\cite{Stelle15PRD}. Here, nevertheless, the trivial flat spacetime is the only solution in the form of Eqs.~\eqref{metric} and~\eqref{FrobeniusAB} that satisfies this condition. This can be verified by comparing the solution~\eqref{Sol00Geral} with the expansion of the equation $R=0$ order by order. The lower-order equations generate constraints between the free parameters of~\eqref{Sol00Geral} and, at order $r^8$, it forces $a_i=0=b_i$ for $i\geqslant 1$.

Also, solutions such that $g_{tt} g_{rr} = -1$ --- i.e., defined by a single metric function, $A(r)=1/B(r)$ --- can be shown to have no free parameters, with all the coefficients in~\eqref{Sol00Geral} being determined by the constants in the action. Simply put, under these conditions the constraints between the free parameters in~\eqref{Sol00Geral} are such that $b_i=0$ for $i=3,4,\ldots$, but it might happen that $b_2$ is a nonzero fixed constant.  Although the assumption $A(r)=1/B(r)$ is often employed in the construction of phenomenological models of regular black holes, our result implies that it might be too restrictive in gravity models with more than four derivatives, in line with statements in~\cite{Bueno:2017sui,Daas:2024pxs}. Indeed, the lack of free parameters makes it difficult to associate solutions of this type with matter sources.
This observation might serve as a motivation for further studies of regular geometries  with $g_{tt} g_{rr} \neq -1$ (see, e.g.,~\cite{NosLW,dePaulaNetto:2023vtg}).

\section{Effect of the terms required by renormalizability}
\label{Sec.Ren}

Up to this point, the discussion concerned the most general (complete) sixth-derivative gravity model. There are various branches, though, that can be analyzed if some of the coefficients in the action~\eqref{mostgeneralaction} are switched off or taken in particular combinations. A more detailed consideration of these specific scenarios will be carried out in a separate work~\cite{Praha2}, but here we would like to address the question of the effect of the terms required by renormalizability.
In fact, while the action~\eqref{mostgeneralaction} emerges in its most general form from quantum corrections to general relativity,  it need not contain all those six-derivative structures if it is taken to be the action of a superrenormalizable gravity.

From the point of view of renormalizability, a sixth-derivative gravity model must have sixth-order derivatives in its spin-2 and spin-0 sectors. In terms of the action~\eqref{mostgeneralaction}, this corresponds to the requirements $\ga_2 \neq 0$ and $3 \ga_1 + \ga_2 \neq 0$. Also, since the counterterms can have up to four metric derivatives~\cite{AsoreyLopezShapiro}, for multiplicative renormalizability we must have $\al,\be_1,\be_2\neq 0$, and a cosmological constant. We omit the latter, for our interest is in the higher derivative's effects. Therefore, the terms in the first line of Eq.~\eqref{mostgeneralaction} with the restrictions mentioned above suffice to yield a superrenormalizable model; omitting or including cubic-curvature structures will not affect the renormalizability.

In this spirit, superrenormalizable models can be formulated in terms of other structures quadratic in curvature and with two covariant derivatives, such as $R_{\mu\nu\al\be} \Box R^{\mu\nu\al\be}$ or $C_{\mu\nu\al\be} \Box C^{\mu\nu\al\be}$, because they also yield sixth-order derivatives of the metric. Although they only differ from combinations of $R\Box R$ and $R_{\mu\nu} \Box R^{\mu\nu}$ by cubic and boundary terms, such models are not equivalent if the action does not contain all the terms present in~\eqref{mostgeneralaction}.

In Table~\ref{Tab2} we summarize the families of solutions that exist around $r=0$ for a superrenormalizable action with all the two- and four-derivative structures and different combinations of six-derivative terms (with arbitrary coefficients). Here we restrict considerations to integer values of $s$ and $t$.

\begin{table}[t]
    \centering
    \begin{tabular}{|c|c|c|}
            \hline
             Six-derivative  & $(s,t)_0$ solution   & Number of free \\ 
             terms in the action		   & family 				& parameters  \\ 
            \hline
            $R\Box R$, $R_{\mu\nu} \Box R^{\mu\nu}$, $R_{\mu\nu\al\be} \Box R^{\mu\nu\al\be}$ & $(0,0)_0$ & $6 \to 5$ \\ 
            \hline
            \multirow{2}{*}{$R\Box R$, $R_{\mu\nu} \Box R^{\mu\nu}$} & \multicolumn{1}{c|}{$(0,0)_0$} & \multicolumn{1}{c|}{ $6 \to 5$ } \\ 
                                 & \multicolumn{1}{c|}{$(1,-1)_0$} & \multicolumn{1}{c|}{ $3 \to 2$ } \\   
            \hline    
            $R \Box R$, $R_{\mu\nu\al\be} \Box R^{\mu\nu\al\be}$ & $(0,0)_0$ & $6 \to 5$ \\ 
            \hline
            \multirow{2}{*}{$R\Box R$, $C_{\mu\nu\al\be} \Box C^{\mu\nu\al\be}$} & \multicolumn{1}{c|}{$(0,0)_0$} & \multicolumn{1}{c|}{ $6 \to 5$ } \\ 
                                 & \multicolumn{1}{c|}{$(2,2)_0$} & \multicolumn{1}{c|}{ $8 \to 7$ } \\ 
            \hline  
        \end{tabular}    
        \caption{\small Summary of solutions around $r=0$. The arrow in the counting of the number of parameters indicates the reduction of parameters after taking into account the freedom to rescale the time coordinate.
         }
\label{Tab2}
\end{table}

Some models admit a family of solutions with indicial structure $(1,-1)_0$ or $(2,2)_0$, in addition to the family $(0,0)_0$. The latter is described as a particular case of~\eqref{Sol00Geral}, by adjusting the coefficients $\ga_i$ according to the terms present in the action. 
Solutions with indicial structures $(0,0)_0$, $(1,-1)_0$, and $(2,2)_0$ also occur in the fourth-derivative gravity, but with a different number of free parameters~\cite{Stelle78,Stelle15PRD}.

The family of solutions $(1,-1)_0$ here is defined by 2 physical parameters, so it is smaller than the one of fourth-derivative gravity, which has 3 parameters~\cite{Stelle15PRD}. This family contains the Schwarzschild solution, which turns out to be the only solution in this class that satisfies $R=0$ (and, besides that, $g_{tt} g_{rr} = -1$). Solutions in this family are singular, with the Kretschmann scalar behaving like $r^{-6}$ as $r \to 0$.

On the other hand, the family $(2,2)_0$ here is larger than in fourth-derivative gravity, for it has two more free parameters. These solutions are also singular, with the Kretschmann scalar behaving like $r^{-8}$ as $r \to 0$. However, they are only present in some models constructed with the Weyl tensor~\cite{endnote1}.

\section{Conclusions}

There is an important difference between fourth- and sixth-derivative gravity in what concerns the space of static spherically symmetric solutions. 
In fact, while the former admits families of singular solutions, with indicial structure $(1,-1)_0$ and $(2,2)_0$~\cite{Stelle78}, in a generic sixth-derivative gravity described by the action~\eqref{mostgeneralaction} (with arbitrary and unrelated couplings) all the solutions of the form of Eqs.~\eqref{metric} and~\eqref{FrobeniusAB} belong to the family $(0,0)_0$ and are regular at $r=0$. 
This possibility was suggested in~\cite{Holdom:2002xy}, and our analysis corroborates the result.
A similar contrast between these higher-derivative gravities had already been noticed for the linear version of the models~\cite{BreTib1}, and it is rewarding to see that 
it
has a counterpart at the nonlinear level.

We have also identified solutions that contain a horizon. The fact that we only found regular solutions around $r=0$ might suggest that these black hole solutions are regular. 
In order to confirm this statement, however, it would be necessary to use numerical methods or to prove that the solutions with a horizon also have a representation in terms of Frobenius series around the origin. 
Indeed, there may still exist singular solutions of non-Frobenius type, or regular solutions at $r=0$ may possess singularities for finite values of $r\neq 0$. 
The study of numerical solutions to sixth-derivative gravity is still in the early stages (see, e.g.,~\cite{Pawlowski:2023dda}), and
we expect that our work will motivate further research on this important topic for the understanding of the role of higher derivatives in gravity. 

Regarding incomplete sixth-derivative models, we discussed the influence of the terms required by renormalizability on the space of solutions. In particular, we identified renormalizable models that have singular solutions of the type $(1,-1)_0$, and others admitting solutions $(2,2)_0$. Nevertheless, since these models are obtained by a special tuning between the coefficients in the general action~\eqref{mostgeneralaction}, we conjecture that the solution that is equivalent to the Schwarzschild one in general relativity is in the class $(0,0)_0$ --- because it is the one common to all such models with sixth-order field equations. The situation here is similar to fourth-derivative gravity, for which the asymptotically flat solution that couples to normal ghost-free matter is in the $(2,2)_0$ class, although the model admits the Schwarzschild black hole as solution~\cite{Stelle15PRD}. 

Last but not least, we found that solutions satisfying the constraints $R=0$ or $g_{tt} g_{rr} = -1$ do not seem to be of much relevance in gravity theories with more than four metric derivatives (with generic couplings). Therefore, like in~\cite{dePaulaNetto:2023vtg}, it might be necessary to relax those conditions to construct phenomenological models of regular black holes aimed at reproducing aspects of solutions of higher-derivative gravity.

\begin{acknowledgements}
We thank Robert \v{S}varc for the fruitful discussions and helpful comments on our work. Both authors acknowledge financial support by Primus grant PRIMUS/23/SCI/005 from Charles University. Additionally, I.K. appreciates the support from the Charles University Research Center Grant No. UNCE24/SCI/016.
\end{acknowledgements}


\appendix*

\section{System of indicial equations and its solution}
\label{App}

The system of indicial equations for the Frobenius series solutions of sixth-derivative gravity around $r=0$ is obtained by substituting the ansatz for the metric, Eqs.~(3) and~(4), into the field equations originated from the action~(1) and expanding to the lowest order in $r$. 
The resulting expressions have the general form of Eq.~(6), but here we shall split the analysis into three cases: $s>0$, $s=0$, and $s<0$.

\subsection{Case $s>0$}

If $s>0$, the expansion of the field equations at lowest order yields
\bea 
\frac{H_{tt}(r)}{B(r)} & = & \frac{1}{32 a_0^3 \, r^{6 + 3s}} \sum_{i=1}^8 \ga_i \, g_i(s,t)   + O \left( \frac{1}{r^{5 + 3s}} \right) ,
\nonumber
\\
H_{rr}(r) &=& \frac{1}{32 a_0^2 \, r^{6 + 2s}} \sum_{i=1}^8 \ga_i \, h_i(s,t) + O \left( \frac{1}{r^{5 + 2s}} \right) ,
\eea
where the coefficients $g_i(s,t)$ and $h_i(s,t)$ depend on the quantities $s$ and $t$ through
\begin{widetext}
{\allowdisplaybreaks
\bea
g_1(s,t) & = & 4 (s+2) [ s (t+4)-t (t+2)-4 ] \big[ (6 s-t) \big(10 s^2-s t-t^2\big) + 2 \big(118 s^2-34 s t-t^2\big) + 4 (69 s-14 t+22) \big] ,
\\
h_1(s,t) & = &  4 (s+2) [ s (t+4)-t (t+2)-4 ] \big[ t \big(10 s^2-s t-t^2\big) +2 \big(20 s^2+6 s t-3 t^2\big)+4 (31 s-6 t+14) \big] ,
\\
g_2(s,t) & = & 2 \big[
s t \big(60 s^4-76 s^3 t+11 s^2 t^2+6 s t^3-t^4\big)
+ 2 \big(60 s^5+146 s^4 t-208 s^3 t^2+35 s^2 t^3+8 s t^4-t^5
+328 s^4+152 s^3 t
\nonumber
\\
&&
-263 s^2 t^2+96 s t^3-3 t^4\big)
+8 \big(124 s^3-69 s^2 t-15 s t^2+21 t^3 -5 s^2-110 s t+2 t^2\big) -96 (9 s+2 t+3)
 \big],
\\
h_2(s,t) & = &  2 \big[
s t^2  \big(10 s^3-11 s^2 t+t^3\big)
+2 t  \big(20 s^4+6 s^3 t-21 s^2 t^2+2 s t^3+t^4\big)
+2  \big(60 s^4+56 s^3 t-71 s^2 t^2-4 s t^3+21 t^4\big)
\nonumber
\\
&&
+8  \big(64 s^3-9 s^2 t-33 s t^2-7 t^3
+    47 s^2-26 s t+6 t^2\big)
+96  (2 t-5 s -3)
\big],
\\
g_3(s,t) & = &  4 [ s (t+4)-t (t+2)-4 ]^2 \big[  30 s^2+s t-t^2 + 2 (47 s-t+37) \big] ,
\\
h_3(s,t) & = &  4 [ s (t+4)-t (t+2)-4 ]^2 [  t  (5 s+t)+  2 (10 s+4 t+23) ] ,
\\
g_4(s,t) & = & 2 \big[ 
t^2 (s-t)^2 \big(30 s^2+s t-t^2\big)
+2 t \big(80 s^4-55 s^3 t-71 s^2 t^2+47 s t^3-t^4\big)
+2 \big(140 s^4+146 s^3 t-193 s^2 t^2+22 s t^3
\nonumber
\\
&&
+25 t^4\big)
+8  \big(59 s^3-20 s^2 t+23 s t^2+12 t^3
-39 s^2+28 s t+56 t^2\big)
-32 (8 s-21 t-19)
\big]  ,
\\
h_4(s,t) & = & 2 \big[ 
t^3  (s-t)^2 (5 s+t)
+8 t^2  \big(5 s^3-6 s^2 t+t^3\big)
+2 t  \big(70 s^3-35 s^2 t-34 s t^2+27 t^3\big)
+8  \big(30 s^3+5 s^2 t-23 s t^2+14 t^3
\nonumber
\\
&&
+ 19 s^2-28 s t+48 t^2\big)
-32  (21 s-16 t -17)
\big],
\\
g_5(s,t) & = &  
t^2  (s-t)^2 \big(30 s^2+s t-t^2\big)
+12 s t  \big(10 s^3-3 s^2 t-14 s t^2+7 t^3\big)
+6  \big(40 s^4+84 s^3 t-23 s^2 t^2-8 s t^3+7 t^4\big)
\nonumber
\\
&&
+8  \big(71 s^3+69 s^2 t+24 s t^2+8 t^3\big)
+24  \big(6 s^2+16 s t+11 t^2\big)
-32 (3 s-9 t-10)
,
\\
h_5(s,t) & = &  
t^3  (s-t)^2 (5 s+t)
+6 t^2  \big(5 s^3-5 s^2 t-s t^2+t^3\big)
+6 t  \big(20 s^3+3 s^2 t-12 s t^2+9 t^3\big)
+8  \big(25 s^3+27 s^2 t-3 s t^2
\nonumber
\\
&&
-5 t^3\big)
+24  \big(10 s^2+8 s t+9 t^2\big)
-32 (9 s-15 t-14)
,
\\
g_6(s,t) & = & 
t^2 (s-t)^2 \big(30 s^2+s t-t^2\big)
+2 t \big(40 s^4+19 s^3 t-97 s^2 t^2+37 s t^3+t^4\big)
+ 4 \big(20 s^4+24 s^3 t-22 s^2 t^2-43 s t^3
\nonumber
\\
&&
+11 t^4\big)
+8  \big(10 s^3-5 s^2 t+23 s t^2-10 t^3
     -25 s^2+54 s t+39 t^2\big)
-32 (5 s-18 t-7) ,
\\
h_6(s,t) & = &  
t^3  (s-t)^2 (5 s+t)
+4 t^2  \big(5 s^3-3 s^2 t-3 s t^2+t^3\big)
+4 s t  \big(10 s^2-9 s t-3 t^2\big)
+8  \big(20 s^3-7 s^2 t-14 s t^2+19 t^3
\nonumber
\\
&&
+ 11 s^2-14 s t+21 t^2\big)
-32 (15 s-4 t -11)
,
\\
g_7(s,t) & = &  4 \big[ 
t^2 (s-t)^2 \big(30 s^2+s t-t^2\big)
+2 t  \big(40 s^4+19 s^3 t-97 s^2 t^2+37 s t^3+t^4\big)
+ 2 \big(40 s^4+128 s^3 t+43 s^2 t^2-92 s t^3
\nonumber
\\
&&
+21 t^4\big)
+16 \big(15 s^3-3 s^2 t+27 s t^2-4 t^3\big)
+8  \big(46 s^2-36 s t+45 t^2\big)
+32 (15 s+13)
\big] ,
\label{f7st}
\\
h_7(s,t) & = &  4 \big[ 
t^3  (s-t)^2 (5 s+t)
+4 t^2  \big(5 s^3-3 s^2 t-3 s t^2+t^3\big)
+2 t  \big(20 s^3+17 s^2 t-20 s t^2+11 t^3\big)
+8  \big(20 s^3-2 s^2 t 
\nonumber
\\
&&
+5 s t^2-t^3
+ 26 s^2-8 s t+31 t^2\big)
-32 (6 s-3 t-11)
\big],
\label{g7st}
\\
g_8(s,t) & = & 4 \big[  
t^2  (s-t)^2 \big(30 s^2+s t-t^2\big)
+6 t^2 \big(31 s^3 -41 s^2 t +9 s t^2 +t^3 +70 s^2-54 s t+4 t^2\big)
-8  \big(s^3-66 s t^2+19 t^3\big)
\nonumber
\\
&&
+32 \big(9 t^2+1\big)
\big],
\label{f9st}
\\
h_8(s,t) & = &  4 \big[  
t^3  (s-t)^2 (5 s+t)
+24 s t^3  (s-t)
+12 t^3  (3 s-t)
+8  \big(10 s^3-3 s^2 t+7 t^3\big)
+144 s^2 
+160
\big].
\label{g9st}
\eea
}
\end{widetext}
Hence, for the field equations to be solved at lowest order in $r$ for \textit{any} values of $\ga_{1,\ldots,8}$, we must have
\beq
\label{ffgghhs>0}
g_i(s,t) = h_i(s,t) =  0, \quad  \forall \, i=1,\ldots,8 ,
\eeq
which constitute the system of indicial equations for $s>0$.

To prove that the system~\eqref{ffgghhs>0} does not admit solutions with $s>0$, consider the following observations:
\begin{itemize}
\item[i.] The subsystem $g_3(s,t) = h_3(s,t) =0$ only admits the one-parameter family of solutions
\beq
\label{onlys}
s = \frac{t^2+2 t+4}{t+4}, \quad t\neq -4.
\eeq
Indeed, it is straightforward to verify that 
\begin{widetext}
\beq
h_3(s,t) = 0 \quad \Longleftrightarrow \quad s = \frac{t^2+2 t+4}{t+4} \quad \text{or} \quad s = - \frac{t^2 + 8 t + 46}{5 (t+4)}.
\eeq
The latter option yields
\beq
g_3(s(t),t) = -\frac{1728 (t-2) (t^2 + 3t +11)^3}{25 (t+4)^2} = 0 , \quad t\in\mathbb{R} \quad \Longleftrightarrow \quad t = 2,
\eeq
\end{widetext}
forcing $s = - 11/5 < 0$, which violates the assumption on the sign of $s$. On the other hand, the former option, given by Eq.~\eqref{onlys}, solves $g_3(s,t) = 0$ for any value of $t\neq -4$.

\item[ii.] The only admissible solution for the subsystem $g_3(s,t) = h_3(s,t) = 0$ and $h_4(s,t) = 0$ is $s = -t = 1$.
\\
In fact, substituting~\eqref{onlys} into the expression for $h_4(s,t)$ we obtain
\beq
h_4(s,t) = \frac{64 (t+1)^2 (t^2 + 3t +11) (t^2 + 4t +12)}{(t+4)^2},
\eeq
whose roots can be easily obtained using Bhaskara's formula. The only real solution is $t=-1$, which yields $s=1$.
\end{itemize} 
Finally, it is immediate to verify that $s = -t = 1$ is not a zero of any of the functions in Eqs.~\eqref{f7st}--\eqref{g9st}. This completes the proof that the system~\eqref{ffgghhs>0} does not have solutions with $s>0$.

\subsection{Case $s=0$}

In this case the expansions of $H_{tt}$ and $H_{rr}$ have the form
\bea
\frac{H_{tt}(r)}{B(r)} &=& \frac{1}{32 a_0^3 \, r^{6}} \sum_{i=1}^8 \ga_i \, g_i(0,t,a_0)   + O \left( \frac{1}{r^{5}} \right) ,
\nonumber
\\
H_{rr}(r) &=& \frac{1}{32 a_0^2 \, r^{6}} \sum_{i=1}^8 \ga_i \, h_i(0,t,a_0) + O \left( \frac{1}{r^{5}} \right) ,
\eea
now with coefficients $g_i(0,t,a_0)$ and $h_i(0,t,a_0)$ that also depend on $a_0$, namely,
\begin{widetext}
{\allowdisplaybreaks
\bea
g_1(0,t,a_0) & = & - 8 \big[4 (1 - a_0) +t (t+2)\big] \big[ 
t^3 
-2 t^2 
-56 t 
+8 (a_0+11)
\big] ,
\label{f1-0t}
\\
h_1(0,t,a_0) & = & 8 \big[4 (1 - a_0) +t (t+2)\big] \big[
t^3 
+6 t^2 
+24 t  
-8 (a_0+7)
\big] ,
\\
g_2(0,t,a_0) & = & - 4 \big[
t^5 
+3 t^4 
-84 t^3 
-8 (a_0+1) t^2 
+96 (a_0+1) t 
-48 \big(a_0^2+2 a_0-3\big)
\big] ,
\\
h_2(0,t,a_0) & = & 4 \big[
t^5 
+21 t^4 
-28 t^3 
+8 (a_0+3) t^2
+32 (3-5 a_0) t 
+48 \big(a_0^2+2 a_0-3\big)
\big],
\\
g_3(0,t,a_0) & = & - 4 \big[4(1- a_0)+t (t+2)\big]^2 \big[
t^2
+2 t
+2 (a_0 - 37)
\big] ,
\\
h_3(0,t,a_0) & = & 4 \big[4 (1 - a_0) +t (t+2)\big]^2 \big[
t^2 
+8 t 
+2 (a_0+23)
\big],
\\
g_4(0,t,a_0) & = & - 2 \big[
t^6 
+2 t^5 
-2 (a_0+25) t^4 
-8 (a_0+12) t^3 
+32 (5 a_0-14) t^2 
+672 (a_0-1) t 
+32 (a_0-1)^2 (a_0-19) 
\big] ,
\\
h_4(0,t,a_0) & = & 2 \big[
t^6 
+8 t^5 
-2 (a_0-27) t^4 
+8 (14-3 a_0) t^3 
+32 (12-7 a_0) t^2 
+32 \big(a_0^2-17 a_0+16\big) t 
+32 (a_0-1)^2 (a_0+17)
\big], \,\,\,\,\,\,\,\,\,\,\,\,
\label{g4-0t}
\\
g_5(0,t,a_0) & = & - \big[
t^6 
-42 t^4 
-64 t^3 
-24 (a_0+11) t^2 
+288 (a_0-1) t 
+32 (a_0-1)^2 (a_0-10) 
\big],
\\
h_5(0,t,a_0) & = &  
t^6 
+6 t^5 
+54 t^4 
-40 t^3 
-24 (a_0-9) t^2 
-480 (a_0-1) t 
+32 (a_0-1)^2 (a_0+14),
\\
g_6(0,t,a_0) & = &  - \big[
t^6 
-2 t^5 
-44 t^4 
-8 (a_0-10) t^3 
+24 (a_0-13) t^2 
+576 (a_0-1) t 
+32 (a_0-1)^2 (a_0-7)
\big],
\\
h_6(0,t,a_0) & = &  
t^6 
+4 t^5 
-8 (a_0-19) t^3 
-168 (a_0-1) t^2 
-128 (a_0-1) t 
+32 (a_0-1)^2 (a_0+11)
\\
g_7(0,t,a_0) & = & - 4 \big[
t^6 
-2 t^5 
-2 (a_0+21) t^4 
+8 (a_0+8) t^3 
+72 (a_0-5) t^2 
+32 (a_0-1)^2 (a_0-13)
\big] ,
\\
h_7(0,t,a_0) & = & 4 \big[
t^6 
+4 t^5 
-2 (a_0-11) t^4 
-8 (a_0+1) t^3 
+8 (31-7 a_0) t^2 
+32 \big(a_0^2-4 a_0+3\big) t 
+32 (a_0-1)^2 (a_0+11)
\big],
\\
g_8(0,t,a_0) & = & - 4 \big[
t^6 
-6 t^5 
-24 t^4 
+152 t^3 
-288 t^2 
+32 (a_0-1)^3
\big],
\\
h_8(0,t,a_0) & = & 4 \big[
t^6 
-12 t^4 
+56 t^3 
+32 (a_0-1)^2 (a_0+5)
\big] .
\label{g9-0t}
\eea
}
\end{widetext}
Therefore, for the field equations to be solved at lowest order in $r$ for \textit{any} values of $\ga_{1,\ldots,8}$, we must have
\beq
\label{ffgghhs=0}
g_i(0,t,a_0) = h_i(0,t,a_0) =  0, \quad  \forall \, i=1,\ldots,8 .
\eeq
Notice that this system of indicial equations for $s=0$ might also act as constraint to the value of the parameter $a_0$ in Eq.~(3).

To prove that the only solution of~\eqref{ffgghhs=0} is $t=0$ and $a_0=1$, we proceed like in the previous case and start by considering the coefficients related to the parameter $\ga_3$. We have the following results:
\begin{itemize}
\item[i.] The only admissible solutions of the subsystem $g_3(0,t,a_0) = h_3(0,t,a_0) = 0$ are either $t=-20$ and $a_0=-143$ or in the form
\beq
\label{onlya0}
a_0 = \frac{1}{4} \left(t^2+2 t+4\right), \quad t\in\mathbb{R} .
\eeq
In fact, it is straightforward to verify that 
\begin{widetext}
\beq
h_3(0,t,a_0) = 0 \quad \Longleftrightarrow \quad a_0 = \frac{1}{4} \left(t^2+2 t+4\right) \quad \text{or} \quad a_0 = -\frac{1}{2} \left(t^2 + 8 t + 46\right).
\eeq
The latter option yields
\beq
g_3(0,t,a_0(t)) = 216 (t+20) (t^2 +6t +32)^2 = 0 , \quad t\in\mathbb{R} \quad \Longleftrightarrow \quad t = -20,
\eeq
\end{widetext}
which results in $a_0 = - 143$. On the other hand, the former solution, Eq.~\eqref{onlya0}, solves $g_3(0,t,a_0) = 0$ for any real $t$.

\item[ii.] The only admissible solution for the subsystem $g_3(0,t,a_0) = h_3(0,t,a_0) = 0$ and $h_4(0,t,a_0) = 0$ is $t=0$ and $a_0=1$.
\\
Indeed, by direct substitution one can verify that $t=-20$ and $a_0=-143$ does not solve $h_4(0,t,a_0)=0$. As for solutions in the form~\eqref{onlya0}, substituting into~\eqref{g4-0t} we obtain
\beq
h_4(0,t,a_0(t)) = 2 t^2 \big(t^2+2\big) (t^2 + 6 t +32),
\eeq
whose only real root is $t=0$, which yields $a_0=1$.
\end{itemize} 
It is immediate to verify that $t=0$ and $a_0=1$ is a zero of all the functions~\eqref{f1-0t}--\eqref{g9-0t}, being, therefore, the only solution of the system~\eqref{ffgghhs=0}.

\subsection{Case $s<0$}

Finally, the lowest-order term of the expansion of $H_{tt}$ and $H_{rr}$ for negative values of $s$ reads
\bea
\frac{H_{tt}(r)}{B(r)} &=& -\frac{ 4 \ga_3 +2 \ga_4 + \ga_5 + \ga_6 +4  \ga_7 +4  \ga_8 }{r^{6}}    + O\big( r^{-5} \big),
\nonumber
\\
H_{rr}(r) &=&  \frac{4 \ga_3 +2 \ga_4 + \ga_5 + \ga_6 +4  \ga_7 +4  \ga_8}{r^{6-s}} a_0     + O\big( r^{s-5} , r^{-6} \big)  .
\nonumber
\\
\eea
Therefore, the field equations cannot be solved at lowest order for arbitrary values of the  parameters $\ga_i$ if $s<0$.

\end{document}